\documentclass[12pt]{article}

\begin{document}

\def\CA{{\cal A}}
\def\CB{{\cal B}}
\def\CD{{\cal D}}
\def\CE{{\cal E}}
\def\CF{{\cal F}}
\def\CG{{\cal G}}
\def\CH{{\cal H}}
\def\CI{{\cal I}}
\def\CJ{{\cal J}}
\def\CK{{\cal K}}
\def\CL{{\cal L}}
\def\CM{{\cal M}}
\def\CN{{\cal N}}
\def\CO{{\cal O}}
\def\CP{{\cal P}}
\def\CQ{{\cal Q}}
\def\CR{{\cal R}}
\def\CS{{\cal S}}
\def\CT{{\cal T}}
\def\CU{{\cal U}}
\def\CV{{\cal V}}
\def\CW{{\cal W}}
\def\CX{{\cal X}}
\def\CY{{\cal Y}}
\def\CZ{{\cal Z}}

\newcommand{\todo}[1]{{\em \small {#1}}\marginpar{$\Longleftarrow$}}
\newcommand{\labell}[1]{\label{#1}\qquad_{#1}} 
\newcommand{\bbibitem}[1]{\bibitem{#1}\marginpar{#1}}
\newcommand{\llabel}[1]{\label{#1}\marginpar{#1}}

\newcommand{\sphere}[0]{{\rm S}^3}
\newcommand{\su}[0]{{\rm SU(2)}}
\newcommand{\so}[0]{{\rm SO(4)}}
\newcommand{\bK}[0]{{\bf K}}
\newcommand{\bL}[0]{{\bf L}}
\newcommand{\bR}[0]{{\bf R}}
\newcommand{\tK}[0]{\tilde{K}}
\newcommand{\tL}[0]{\bar{L}}
\newcommand{\tR}[0]{\tilde{R}}

\newcommand{\btzm}[0]{BTZ$_{\rm M}$}
\newcommand{\ads}[1]{{\rm AdS}_{#1}}
\newcommand{\ds}[1]{{\rm dS}_{#1}}
\newcommand{\eds}[1]{{\rm EdS}_{#1}}
\newcommand{\sph}[1]{{\rm S}^{#1}}
\newcommand{\gn}[0]{G_N}
\newcommand{\SL}[0]{{\rm SL}(2,R)}
\newcommand{\cosm}[0]{R}
\newcommand{\hdim}[0]{\bar{h}}
\newcommand{\bw}[0]{\bar{w}}
\newcommand{\bz}[0]{\bar{z}}
\newcommand{\beq}{\begin{equation}}
\newcommand{\eeq}{\end{equation}}
\newcommand{\beqn}{\begin{eqnarray}}
\newcommand{\eeqn}{\end{eqnarray}}
\newcommand{\pat}{\partial}
\newcommand{\lp}{\lambda_+}
\newcommand{\bx}{ {\bf x}}
\newcommand{\bk}{{\bf k}}
\newcommand{\bb}{{\bf b}}
\newcommand{\BB}{{\bf B}}
\newcommand{\tp}{\tilde{\phi}}
\hyphenation{Min-kow-ski}

\newcommand{\beql}[1]{\begin{equation}\label{eq:#1}}
\newcommand{\eq}[1]{(\ref{eq:#1})}
\newcommand{\ack}[1]{{\bf Pfft!: #1}}\newcommand{\talkpt}[1]{{\bf TP: #1}}
\newcommand{\field}[1]{\mathbb{#1}}
\newcommand{\CC}{{\field C}}
\newcommand{\RR}{{\field R}}
\newcommand{\ZZ}{{\field Z}}

\newcommand{\scale}{\sigma}

\def\apr{\alpha'}
\def\str{{str}}
\def\lstr{\ell_\str}
\def\gstr{g_\str}
\def\Mstr{M_\str}
\def\lpl{\ell_{pl}}
\def\Mpl{M_{pl}}
\def\varep{\varepsilon}
\def\del{\nabla}
\def\grad{\nabla}
\def\tr{\hbox{tr}}
\def\perp{\bot}
\def\half{\frac{1}{2}}
\def\p{\partial}
\def\perp{\bot}
\def\eps{\epsilon}

\def\NPB{{\it Nucl. Phys. }{\bf B}}
\def\PL{{\it Phys. Lett. }}
\def\PRL{{\it Phys. Rev. Lett. }}
\def\PRD{{\it Phys. Rev. }{\bf D}}
\def\CQG{{\it Class. Quantum Grav. }}
\def\JMP{{\it J. Math. Phys. }}
\def\SJNP{{\it Sov. J. Nucl. Phys. }}
\def\SPJ{{\it Sov. Phys. J. }}
\def\JETPL{{\it JETP Lett. }}
\def\TMP{{\it Theor. Math. Phys. }}
\def\IJMPA{{\it Int. J. Mod. Phys. }{\bf A}}
\def\MPL{{\it Mod. Phys. Lett. }}
\def\CMP{{\it Commun. Math. Phys. }}
\def\AP{{\it Ann. Phys. }}
\def\PR{{\it Phys. Rep. }}

\renewcommand{\thepage}{\arabic{page}}
\setcounter{page}{1}

\rightline{hep-th/0302230}
\rightline{VPI-IPPAP-03-02}
\rightline{CERN-TH/2003-xxx}
\vskip 1cm
\centerline{\Large \bf Deconstruction and Holography}\vskip0.25cm
\vskip 1cm
\vskip 1cm

\renewcommand{\thefootnote}{\fnsymbol{footnote}}
\centerline{{\bf Vishnu Jejjala,${}^{1}$\footnote{vishnu@vt.edu}
Robert G. Leigh,${}^{2,3}$\footnote{rgleigh@uiuc.edu} and
Djordje Minic${}^{1}$\footnote{dminic@vt.edu}
}}
\vskip .5cm
\centerline{${}^1$\it Institute for Particle Physics and Astrophysics}
\centerline{\it Department of Physics, Virginia Tech}
\centerline{\it Blacksburg, VA 24061, U.S.A.}
\vskip .5cm
\centerline{${}^2$\it CERN-Theory Division}
\centerline{\it CH-1211, Geneva 23, Switzerland}
\vskip .5cm
\centerline{${}^3$\it Department of Physics}
\centerline{\it University of Illinois at Urbana-Champaign}
\centerline{\it 1110 W. Green Street, Urbana, IL 61801, U.S.A.}

\vskip .5cm

\setcounter{footnote}{0}
\renewcommand{\thefootnote}{\arabic{footnote}}

\begin{abstract}
It was recently pointed out that the physics
of a single discrete gravitational extra dimension
exhibits a peculiar UV/IR connection relating 
the UV scale to the radius of the effective extra dimension.
Here we note that this non-locality is a manifestation 
of holography, encoding the correct scaling of the number
of fundamental degrees of freedom of the UV
theory. This in turn relates the Wilsonian
RG flow in the UV theory to the effective 
gravitational dynamics in the extra dimension.
The relevant holographic c-function is determined
by the expression for the holographic bound.
Holography in this context is a result of the requirements of 
unitarity and diffeomorphism invariance.
We comment on the relevance of this observation
for the cosmological constant problem.

\end{abstract}
\newpage

\section{Introduction and relation to previous work}

Recently, we have argued that the vanishing of the vacuum energy
of $(2+1)$-dimensional gravity \cite{wittencc,Hen,wittencc2,bbs}
may be deconstructed \cite{nima} to $3+1$ dimensions under certain
conditions \cite{ljm}. Our discussion pointed
towards a possibility that there exists a well-defined UV completion 
of $(3+1)$-dimensional gravity. (For related discussions, see 
Refs.\ \cite{others} and \cite{harvard}.)

More recently it was pointed out by
Arkani-Hamed and Schwartz \cite{nimas} that
the physics of gravitational deconstruction
exhibits a fascinating relation between
the characteristic UV and IR scales.
This UV/IR relation was argued \cite{nimas}
to indicate the presence of non-local interactions
in the defining UV theory.

Here, we point out that the UV/IR relation
found in Ref.\ \cite{nimas} is nothing but a manifestation
of the holographic principle \cite{holography,holography1,bhent}.
In particular the specific form of the UV/IR correspondence
found in the problem of deconstruction of gravity
is very analogous to a similar relation
found \cite{susswitt} in a seemingly completely unrelated 
topic - the AdS/CFT correspondence \cite{adscft} (see also \cite{stromds})!

As reviewed in Ref.\ \cite{ljm},
in $(2+1)$-dimensional theories it is possible \cite{wittencc} to have 
vanishing vacuum energy in the absence of a mass degenerate spectrum
of bosonic and fermionic states. The basic point is that
the vacuum state is supersymmetric, 
but the excited states are not mass degenerate because unbroken 
global supercharges do not exist in $2+1$ dimensions \cite{Hen}. 
Any excited state gives a conical geometry whose deficit angle
prohibits spinor fields with covariantly constant asymptotics.  Thus,
there are no global supercharges and no mass degeneracy between Bose
and Fermi excitations.  The size of the non-degeneracy
of the spectrum of low-energy excitations scales as the inverse power
of the three-dimensional Newton constant under the assumption of weak
gravitational coupling \cite{wittencc2}.

The idea behind deconstruction \cite{nima} is that the UV region of a 
theory might
be described in terms of a co-dimension one theory.
Motivated by this idea in Ref.\ \cite{ljm} we argued that 
the $(2+1)$-dimensional phenomenon 
described above can be deconstructed as follows:

\begin{itemize}
\item (1) Assume a local spatial foliation of 4d spacetime.

\item (2) Deconstruct the vacuum part of pure 4d gravity from 
($N$ copies of) 3d general relativity \cite{wittencs}
coupled to certain 3d 
matter fields represented in terms of currents. Assume that the 4d 
sources can be defined in terms of a deconstructed 3d 
theory. 
For sources represented by gauge fields this should be possible 
given the discussion in Ref.\ \cite{ljm}.

\item (3) In the deep UV we have ($N$ copies of) 3d gravity coupled to some
3d sources.  Whatever the matter content of this
3d theory is, the resulting geometry has to
be conical.  Thus, provided we have 3d (but not 4d!) supersymmetry, 
Witten's argument applies: the vacuum is supersymmetric,
yet the excited states are not.

\item (4) In the range of intermediate scales, we have $N$ linked copies
of 3d gravity coupled to 3d currents. Once again, the resulting 3d 
geometry is conical. Thus Witten's argument holds in
the region between the UV and IR.
\end{itemize}

Note that on dimensional grounds, 
the mass splitting should be 
inversely proportional to the three-dimensional Newton constant and 
should vanish at zero deficit angle.
Thus as long as the three-dimensional Newton constant
is of order one as the continuum limit is taken,
and the deficit angle (on each local three-dimensional
slice) is taken to scale as the inverse of the lattice spacing,
the Bose-Fermi splitting will be finite in the infrared.

According to the outlined argument the vacuum energy is zero in the UV,
and also some place in between UV and IR.  But does it remain zero in the IR?
In the concluding part of this note, we remark that the UV/IR relation
found in Ref.\ \cite{nimas}, which as we argue is just
the statement of the saturation of the holographic bound, can
be utilized to put an upper bound on the maximum value of the
deconstructed cosmological constant.

\section{Holography and Deconstruction}

Before we proceed to establish a relation between the UV/IR correspondence
found in Ref.\ \cite{nimas} and the holographic bound it is useful to 
remember how holography arises in the framework of our previous paper
\cite{ljm}.

The Bekenstein-Hawking bound on
entropy \cite{bhent} arises as follows.  Let us first suppose that the $(2+1)$-dimensional matter
fields are local.  The coupling of $(2+1)$-dimensional gravity to matter is of the 
general form
\begin{equation}
S_{EH} = \frac{1}{G_3} \int d^3x\, \sqrt{-g^{(3)}}\, (R^{(3)} + {\cal L}_{\rm matter}).
\end{equation}
The entropy of local matter degrees of freedom scales as the two-dimensional 
area.  As there are $N$ copies, we have
\begin{equation}
{\cal{S}} \propto \frac{NA}{G_3}.
\end{equation}
Obviously this expression does not have the correct mass dimension.  
The crucial step at this point is to remember that the usual
prescription for dimensional reduction determines the pre-factor to
be $1/G_3 L$, where $L=Na$ is the size of the fourth (lattice) dimension.
Thus, on dimensional grounds,
\begin{equation}
{\cal{S}} \simeq \frac{NA}{G_3 L} = \frac{A}{G_3 a} = \frac{A}{G_4}, 
\end{equation}
which reproduces the Bekenstein-Hawking scaling in $3+1$ dimensions.

Note that this reasoning applies only in the case of 3d to 4d, 
where there are no local gravitational degrees of freedom.
We wish now to relate this scaling to the UV/IR relation found
in Ref.\ \cite{nimas}. The latter result is stronger, applying in any dimension.

\section{Deconstruction and UV/IR mixing}

As argued by Arkani-Hamed and Schwartz in a recent paper \cite{nimas},
there exists a subtlety in the implementation of deconstruction in the 
gravitational context. Namely, the cutoff of the theory cannot be taken to
be $M_4$, the four-dimensional Planck scale, as one would expect. 
A simple tree-level calculation indicates that instead, 
there are amplitudes involving longitudinal components of gravitons which
de-unitarize around a scale
\begin{equation}
\Lambda\sim \left(\frac{M_4^2}{L^5a^2}\right)^{1/9},
\end{equation}
where $L=Na$, $a$ being the lattice spacing, and $G_4 = G_3 a$,
as in Ref.\ \cite{ljm}.
This result seems to be indicative of a UV/IR mixing phenomenon. Indeed, 
when one looks more closely at the details of the effective action for 
the interactions, one finds that one may indeed interpret it as non-local.

In fact, if we require that the cutoff be above the most massive Kaluza-Klein states but
below the unitary threshold, one finds that the highest cutoff that the 
theory may possess is of order
\begin{equation}\label{eq:mcut}
\Lambda_m  \sim \left(\frac{M_4^2}{L}\right)^{1/3}.
\end{equation}
This scale has an important implication in terms of holography. To demonstrate
this, consider a calculation of entropy. In the four-dimensional theory, we 
would estimate
\begin{equation}
{\cal{S}} \sim  A L \Lambda^3,
\end{equation}
which would give the standard wrong result if $\Lambda\sim M_4$.
With the cutoff of eq.\ (\ref{eq:mcut}), we find instead\footnote{Here we
used the thermodynamic relation ${\cal S} \sim V T^3$ where the volume
$V\sim A L$ and $T \sim \Lambda_{m}$, by construction.}
\begin{equation}
{\cal{S}} \sim A L \Lambda_m^3 \sim \frac{N A}{G_3 L} 
\sim \frac{A}{G_4},
\end{equation}
which is nothing but the holographic bound on the
number of degrees of freedom in the UV theory,
{\em as it should be if the theory is really expected
to describe a UV definition of gravitational 4d
dynamics.} The deconstructed theory resists the temptation to lift its 
cutoff too high. It appears that unitarity plus diffeomorphism 
invariance are sufficient to imply holography!
As indicated above, this argument generalizes to any number of dimensions.

As an aside, it is interesting to observe that the above expression
for the holographic bound is reminiscent of the one obtained
from the heuristic
argument  
based on the properties of gravitational focusing \cite{holography1}. 
Given a boundary
region of area $A$, the number of UV degrees of freedom is
estimated from the effective volume determined by the area and
the Planck length $A L_P$ and the UV cut-off given by $1/L_P^3$
which combines into the holographic (Bekenstein-Hawking) 
bound $A/L_P^2$.\footnote{We thank Nemanja Kaloper for an enjoyable 
discussion concerning these heuristics.}

Notice that this kind of relation between a UV/IR correspondence
and holography is in complete analogy to what happens
in the context of the AdS/CFT correspondence, even though
the two topics seem unrelated.
Taking clues from the AdS/CFT correspondence,
we also see that the UV/IR mixing found in the context of
gravitational deconstruction
can be interpreted locally. That is, the local rescaling in the UV theory 
corresponds to the rescaling of the size of the extra dimension:
the UV Wilsonian evolution corresponds to 
the gravitational evolution in the extra dimension.

More explicitly, a local form of the UV/IR correspondence
can be recast in the form of a holographic RG formalism even in the
present discussion of the deconstruction of gravity.

This formalism runs as follows~\cite{holorg,holorg1}. First we 
fix the gauge so that the bulk metric can be written as
\begin{equation}
ds^2 = dr^2 + g_{ij}dx^i dx^j.
\end{equation}
This is just the ADM gauge discussed both in Ref.\ \cite{ljm,nimas}:
the shift vector is set to zero and
the lapse to one.
As noticed above, the UV rescaling corresponds to
the rescaling in the size of the extra dimension, which in the chosen
gauge is nothing but the natural evolution parameter.
Given the fact that the $(3+1)$-dimensional gravity theory is
reparametrization invariant, the local UV rescaling
is encapsulated in the IR by the four-dimensional Hamiltonian constraint
\begin{equation}
{\cal{H}} =0.
\end{equation}
More explicitly
\begin{equation}
{\cal{H}}= (\pi^{ij} \pi_{ij} - \frac{1}{2} \pi^{i}_{i} \pi^{j}_{j})
+\frac{1}{2} \pi_{I} G^{IJ}\pi_{J} + {\cal{L}}.
\end{equation}
Here $\pi_{ij}$ and $\pi_{I}$ are the canonical momenta conjugate to
$g^{ij}$ and $\phi^I$ 
\begin{equation}
\pi_{ij} = \frac{1}{\sqrt{-g}}\frac{\delta S}{\delta g^{ij}}, \quad
\pi_I = \frac{1}{\sqrt{-g}}\frac{\delta S}{\delta \phi^{I}}.
\end{equation}
Here $\phi^I$ denotes some background matter
fields coupled to $(3+1)$-dimensional gravity --- 
for example, the Standard model fields; 
${\cal{L}}$ is a local Lagrangian density, and
$G^{IJ}$ denotes the metric on the space of background matter fields.

As in the context of the AdS/CFT duality~\cite{holorg,holorg1}, the Hamiltonian
constraint can be formally rewritten as a renormalization group equation for
the dual RG flow.\footnote{Here
we follow the formalism of Ref.\ \cite{holorg}.}
In the Hamiltonian constraint
\begin{equation}
\frac{1}{\sqrt{-g}}\left( \frac{1}{2} \left(g^{ij}
{\frac{\delta S}{\delta g^{ij}}}\right)^2
-{\frac{\delta S}{\delta g^{ij}}}{\frac{\delta S}{\delta g_{ij}}}
-\frac{1}{2} G^{IJ}
\frac{\delta S}{\delta \phi^{I}} \frac{\delta S}{\delta \phi^{I}}\right)
= \sqrt{-g}\, {\cal{L}},
\end{equation}
assume that the local four-dimensional action $S$ can be separated into a
local and a non-local piece
\begin{equation}
S(g, \phi) = S_{loc}(g, \phi) + \Gamma (g, \phi).
\end{equation}
Given this rewriting of the four-dimensional action, the Hamiltonian constraint 
can be formally rewritten as a Callan-Symanzik renormalization group 
equation for the effective action~\cite{holorg} $\Gamma$ of the UV
theory at the scale $\Lambda$
\begin{equation}
\frac{1}{\sqrt{-g}}\left( g^{ij}
{\frac{\delta }{\delta g^{ij}}} - \beta^I \frac{\delta}{\delta
\phi^I}\right) \Gamma = HO,
\end{equation}
where $HO$ denotes higher derivative terms of the expression for the
four-dimensional conformal anomaly.
Here the ``beta-function" is defined (in analogy with the
AdS situation) to be $\beta^I = \partial_{\Lambda}
\phi^{I}$,
where $\Lambda$ denotes the cut-off of the defining UV theory.

In the context of the holographic RG formalism developed in the
AdS/CFT correspondence, it is also possible to introduce a 
holographic ``c-function'' which measures the number of accessible 
degrees of freedom and which decreases during RG flow. When the 
spacetime is four-dimensional, one has \cite{holorg,holorg1}
\begin{equation}
c \sim \frac{1}{G \theta^2},
\end{equation}
where $\theta$ is the trace of the extrinsic curvature of the 
boundary surface.\footnote{The trace of the quasi-local Brown-York 
stress \cite{by} tensor turns out to be
$\langle T^{i}_{i} \rangle \sim \theta$,
up to some terms constructed from local intrinsic curvature invariants
of the boundary.    Therefore the RG equation of the defining UV theory is 
given by $\langle T^{i}_{i} \rangle = \beta^I \frac{\partial \Gamma}
{\partial \phi^I}$.}
In the context of the AdS/CFT correspondence the Raychauduri equation, that is,
gravitational focusing, implies monotonicity of the holographic ``c-function''
\begin{equation}
\frac{d \theta}{dt} \le 0,
\end{equation}
as long as a form of the weak positive energy condition is satisfied
by the background test matter fields.  

The important point here is that the holographic ``c-function'' 
is determined by the holographic bound, that is
the Bekenstein-Hawking entropy.
In our context the Bekenstein-Hawking entropy is determined by
$A L \Lambda_{m}^3$.
Thus at a scale $\Lambda$ below the maximal scale,
the natural expression for the holographic ``c-function''
in the present context is precisely the quantity
\begin{equation}
c \sim A L \Lambda^3
\end{equation}
which measures the number of degrees of freedom in the UV theory.

\section{Conclusion: UV/IR, cosmological constant, non-local interactions
and all that}

The argument for the vanishing of the cosmological
constant in $(3+1)$-dimensional gravity as presented in Ref.\ \cite{ljm} is obscured 
by the region of strong coupling in the infrared.
The question is whether the cosmological constant remains zero all
the way at long distances even in the presence of strongly coupled physics.
In the conclusion of this note, we argue that one can actually derive the 
upper bound on the
value of the induced four-dimensional vacuum energy based
on the above discussion regarding the relation deconstruction
and holography.

The essential point is that the picture of vacuum energy based on
deconstruction \cite{ljm} naturally presents us with two different 
energy scales in the IR.
One scale is the mass splitting between fermion and bosonic degrees of
freedom $M_s$, indicating the crucial importance of supersymmetry in our
argument \cite{ljm}, and the other is the four-dimensional Planck scale $M_P$.
These two scales come from the two dimensionful parameters provided
by the UV definition of the infrared physics via deconstruction:
the lattice spacing $a$ and the three-dimensional Newton's constant
$G_3$ \cite{ljm}.

Now, given these two mass scales and the requirement that in the limit
when the mass splitting of the bosonic and fermionic degrees of freedom
goes to zero,
the four-dimensional vacuum energy should go to zero as well,
we can associate one four-dimensional scale $m$ with
$M_s$ and $M_P$.
This scale will provide the natural cut-off in the computation
of the four-dimensional vacuum energy.

Dimensional analysis and the requirement that
$m \to 0$ when $M_s \to 0$ dictates that
\begin{equation}
m \sim \frac{M_s^2}{M_P}.
\end{equation}
This, we claim, is the only effective UV scale left in the problem
in four dimensions.
Notice that this relation is also a manifestation of a UV/IR correspondence.
The $M_P$ is already an IR scale from the point of view
of the defining UV theory. That follows from the relation
$G_4 = G_3 a$. As we approach the continuum, the three-dimensional 
scale is much higher than the effective four-dimensional gravitational scale.
Now, given the fact that the effective action contains
fields coupled to four-dimensional gravity, one expects that the natural
cut-off scale $m$ goes as an inverse power of the IR scale, which
is set by the Planck scale $M_P$, and the scale $M_s$ that governs the Bose-Fermion mass splitting.
The quadratic scaling of $m$ with $M_s$ is determined by
the gravitational coupling of the matter fields at the scale $M_s$.
(By the deconstruction of Witten's argument \cite{ljm}, the vacuum
energy is still zero at this scale.)
Then the above formula indeed follows by dimensional analysis.

When evaluating the vacuum diagrams in order to estimate
the upper bound on the vacuum energy in the infrared we should 
therefore use $m$ as
the only effective cut-off scale.
The naive expression for the vacuum energy is bounded by $m^4$ or 
\begin{equation}
\lambda \sim M_P^4 \left(\frac{M_s}{M_P}\right)^8,
\end{equation}
which is a formula previously discussed in
the literature \cite{banks}.
Therefore, provided the large ratio of the mass splitting to
the Planck scale we get the observed bound on the vacuum energy
density!

Note that this argument is based on dimensional analysis,
the UV/IR relation discussed above, and the fact that
the deconstruction of Witten's argument for the vanishing
of the cosmological constant in three dimensions implies 
zero vacuum energy at a very low scale, set by the value of $M_s$.
The violation of the usual effective field theory reasoning
comes from the UV/IR relation and the vanishing of the
cosmological constant at the scale determined by $M_s$, as implied
by the deconstruction of Witten's argument.

Of course, we have not presented a detailed calculation, and it is not
completely clear if $m^4$ really determines the cosmological constant.
Also, one should carefully consider radiative corrections.
Given the fact that the cosmological constant vanishes at
the scale determined by $M_s$ by deconstruction, 
the radiative corrections determined
by usual effective field theory, cannot be expected to be very large.
It is desirable therefore to provide an explicit calculation to
show this.

In concluding, we raise the question of whether the non-local
interaction expected in the UV theory \cite{nimas} should be of
the type discussed in the ``wormhole'' program \cite{wormhole}.
The idea here is that a guiding principle for the
construction of a defining UV theory should be
the recovery of the full path integral over four-dimensional metrics
in the IR.
This, at least at the level of Euclidean gravity, can be
implemented by an insertion of bilocal operators which
create ``wormholes''. It was shown in the late 80's 
\cite{wormhole} that such topology changing processes
make all couplings in the Wilsonian effective action
describing the interaction of gravity and matter into
true dynamical random variables. It would be interesting to
see whether there is a natural implementation of this idea in
the present context.

\section*{Acknowledgments}
\noindent We thank Vijay Balasubramanian and Nemanja Kaloper for conversations and comments.
VJ thanks the High Energy Group at the University of Pennsylvania for their kind hospitality.  
RGL thanks the Instituto de F\'{\i}sica Te\'orica de la Universidad Aut\'onoma de Madrid for their kind hospitality.
DM thanks the Cosmology Group of the University of California at Davis for their kind hospitality.
This work is supported in part by the U.S.\ Department of Energy under contracts DE-FG02-91ER40677 and DE-FG05-92ER40709.


\end{document}